# Hanbury Brown-Twiss effect and thermal light ghost imaging

Li-Gang Wang,[1,2] Sajid Qamar,[3] Shi-Yao Zhu [1, 2] and M. Suhail Zubairy [3,4]

[1]*Department of Physics, Zhejiang University 310027, Hangzhou, China*

[2]*Department of Physics, Chinese University of Hong Kong, Shatin, N. T., Hong Kong*

[3] *Center for Quantum Physics, COMSATS Institute of Information Technology, Islamabad, Pakistan*

[4]*Institute for Quantum Studies and Department of Physics, Texas A&M University, College Station, Texas 77845, USA*

## Abstract

We show that the essential physics of the Hanbury Brown-Twiss (HBT) and the thermal light ghost imaging experiments is the same, i.e., due to the intensity fluctuations of the thermal light. However, in the ghost imaging experiments, a large number of bits information needs to be treated together, whereas in the HBT there is only one bit information required to be obtained. In the HBT experiment far field is used for the purpose of easy detection, while in the ghost image experiment near (or not-far) field is used for good quality image.

Recently, there is a heated discussion on the physics of the ghost imaging (GI) with thermal light [1-5]. For example, in a recent paper Scarcelli, Berardi and Shih pose the question: 'Can two-photon correlation of chaotic light be considered as correlation of intensity fluctuations?' [5]. The authors conclude that the chaotic (thermal) light ghost image could not be explained with classical mechanics, and the physics of the ghost imaging is not the intensity fluctuations of the thermal light $\langle \Delta I_1 \Delta I_2 \rangle = \langle I_1 I_2 \rangle - \langle I_1 \rangle \langle I_2 \rangle$ as in the Hanbury Brown-Twiss (HBT) experiment [2,5]. Instead, they claim, that the essential physics is the two photon quantum interference [2,5]. On the other hand, others have concluded that the ghost imaging with classical thermal field is essentially a classical effect [3,4].

Originally the ghost image was achieved with entangled light [6]. In 2004 the formation of ghost image with thermal light was predicted [7] and the equation for the image formation is given in Ref. [8]. In 2005 the experiments on ghost imaging with thermal light were realized [9-11]. Since then, theoretical models are put forward to explain the thermal light ghost imaging [3,12, 13]. Until today, far field is used in the HBT experiments, while near field (not far field) is used in the ghost imaging experiments [2,5]. There is a first order coherence for the far field, while the first order coherence for the near field (not far field) is small. A question of interest is whether the far field and the near field result in significantly different physics. We ask ourselves, what will be the results if we use the far field and the near field for both HBT and GI experiments. In the present paper, we address this question and discuss how an answer to this question reveals the physics behind the GI and the HBT effect.

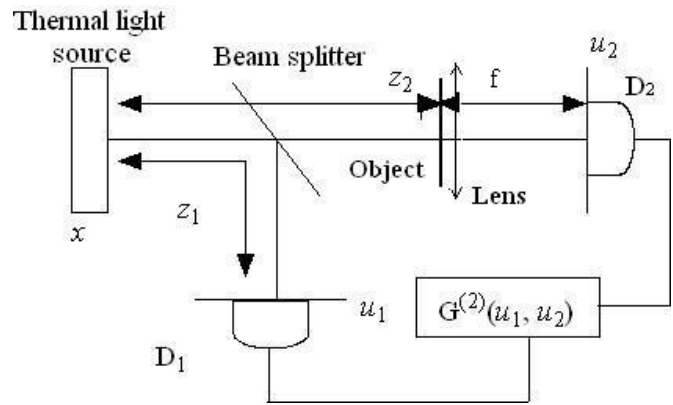

Fig.1 Ghost image setup.

The setup for the thermal light ghost imaging experiment is presented in Fig. 1 with a lens immediately behind the object focusing onto $D_2$ (as a bucket detector [5]). The HBT experiment [14] is the same except the detector $D_2$ is placed at the location of the object and there are no object and lens. In both experiments, the source is a surface thermal light, for example a black box at a certain temperature. The thermal light from the source is split by a beam splitter and shines the two detectors, $D_1$ and $D_2$ through two paths, as shown in Fig.1. The second order correlation is detected by $D_1$ and $D_2$ in both the experiments. For the thermal light, the field statistics is Gaussian and this allows the calculation of the second order correlation from the first order correlation, i.e.,

$$G^2(u_1, u_2) = \langle I_1 I_2 \rangle = <I(u_1)><I(u_2)> + |\Gamma(u_1, u_2)|^2, \quad (1)$$

where $<I(u_{1,2})>$ are the intensities at the points $u_1$ and $u_2$ on the planes of $D_1$ and $D_2$, and $\Gamma(u_1, u_2) = <E^+(u_1)E(u_2)>$ is the cross correlation. The correlations at the detectors can be found out from the correlation at the source by correlation propagation method [14].

$$I(u_{1,2}) = <E^+(u_{1,2})E(u_{1,2})> = \iint <E_s^+(x_1)E_s(x_2)> h_{1,2}^*(x_1, u_{1,2})h_{1,2}(x_2, u_{1,2})dx_1 dx_2, \quad (2)$$

$$\Gamma(u_1, u_2) = <E^+(u_1)E(u_2)> = \iint <E_s^+(x_1)E_s(x_2)> h_1^*(x_1, u_1)h_2(x_2, u_2)dx_1 dx_2, \quad (3)$$

where $x_{1,2}$ are the points at the source plane and the integrals are within the source. Here $h_{1,2}(x, u_{1,2})$ are the propagation functions of the correlation from the source to the detectors alone path 1 and 2, respectively, which depend on the optical elements in the paths. For path 1, $h_1(x, u_1)$ is the same in the two experiments.

$$h_1^{H,G}(x, u_1) = (-i/\lambda z_1)^{1/2} \exp\left[-\frac{i\pi}{\lambda z_1}(x^2 - 2x u_1 + u_1^2)\right]. \quad (4)$$

For path 2, we have different $h_2(x, u_2)$ for the two experiments,

$$h_2^H(x, u_2) = (-i/\lambda z_2)^{1/2} \exp\left[-\frac{i\pi}{\lambda z_2}(x^2 - 2x u_2 + u_2^2)\right], \quad (5a)$$

$$h_2^G(x, u_2) = (-\frac{i}{\lambda f})^{1/2}(-\frac{i}{\lambda z_2})^{1/2} \int dv H(v) \exp\left[-\frac{i\pi}{\lambda z_2}(x^2 - 2x v + v^2) - \frac{i\pi}{\lambda f}(-2vu_2 + u_2^2)\right], \quad (5b)$$

where the superscripts H and G indicate HBT and GI, respectively. It is implicit that $h_{1,2}^{H,G}(x, u_{1,2})$ depend also on $z_1$ or $z_2$. In Eq. (5b), $H(v)$ is the transmittance of the object and the integration is due to the bucket detector. In general, we should consider the two-dimensional imaging. However as $x$ and $y$ directions are independent, we only consider the $x$ direction (one dimension). This however does not affect the physics.

For the thermal light at the source, the first order correlation can be written as a series of the form [15,16]

$$<E_s^+(x_1)E_s(x_2)> \propto 1 - \alpha(x_1 - x_2)^2 + \beta(x_1 - x_2)^4 + \cdots \quad (6)$$

with $\beta/\alpha = 2.2$. This however does not allow for an analytical solution for the correlation functions of the field. We therefore approximate the first order correlation function of the source to be a Gaussian Schell model source [15,17]

$$< E_s^+(x_1)E_s(x_2) >= G_0 \exp\left[-\frac{x_1^2 + x_2^2}{4\sigma_I^2} - \frac{(x_1 - x_2)^2}{2\sigma_g^2}\right]. \quad (7)$$

Here we have a Gaussian distribution for the intensity of the source with the width $\sigma_I$ and $\sigma_g$ is the first order transverse coherence width (correlation length) of the thermal light source. The normalized second order correlations (HBT or GI) for the two experiments are

$$\text{HBT or GI}(u_1, u_2, z_1, z_2) = \frac{|\Gamma(u_1, u_2)|^2}{(<I(u_1)><I(u_2)>)}. \quad (8)$$

First we consider the HBT experiment. We consider point detectors located at $u_1 = 0$ and $u_2 = 0$. Setting $u_1 = u_2 = 0$ in Eq. (8) and substituting Eqs. (2), (3) and (7) into Eq. (8), we obtain

$$HBT(z_1, z_2) = \left\{\frac{A(\bar{z}_1^2)A(\bar{z}_2^2)}{A^2(\bar{z}_1\bar{z}_2) + 16\pi^2(1/\bar{\sigma}_I^2 + 2/\bar{\sigma}_g^2)^2[(\bar{z}_1 - \bar{z}_2)/\bar{z}_1\bar{z}_2]^2}\right\}^{1/2}, \quad (9)$$

where $A(x) = 16\pi^2/x + [(1/\bar{\sigma}_I^2 + 4/\bar{\sigma}_g^2)/\bar{\sigma}_I^2]$, and $\bar{z}_{1,2}, \bar{\sigma}_{I,g}$ are in unit of $\lambda$. If $(\bar{z}_1 - \bar{z}_2) = 0$, we have HBT = 1. As $(\bar{z}_1 - \bar{z}_2)$ increases, HBT decreases. In Fig.2, we plot HBT versus $(\bar{z}_1 - \bar{z}_2)$ with $\bar{\sigma}_g = 10$ and $\bar{z}_1 = 10^6$ for $\bar{\sigma}_I/\bar{z}_1 = 0.002, 0.005, 0.01,$ and $0.1$. In the following, we define $\bar{\sigma}_I/\bar{z}_1 < 0.05$ as the far field, and otherwise as the near field.

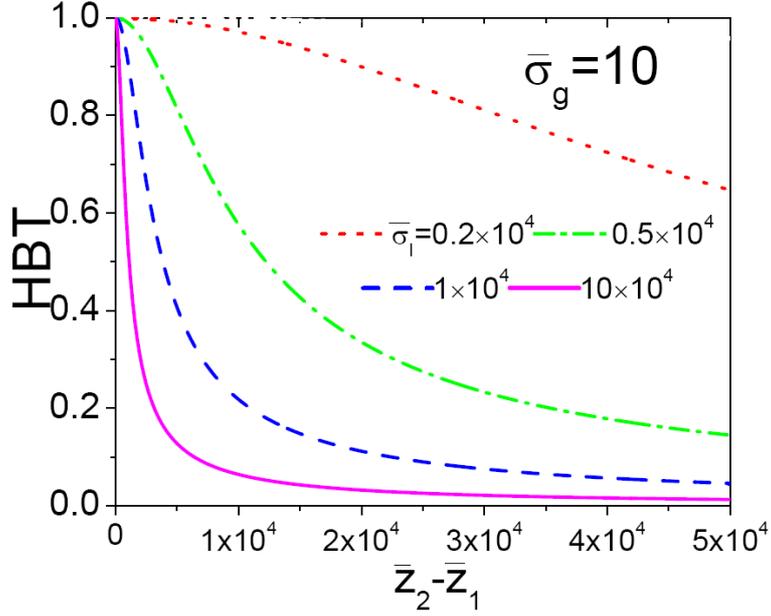

Fig. 2 HBT effect for different $\bar{\sigma}_I / \bar{z}_1 =$ 0.002, 0.005, 0.01, 0.1.

In the experiment with thermal light, the condition $\bar{z}_{1,2}, \bar{\sigma}_I{}^2 \gg \bar{\sigma}_g{}^2$ is always satisfied. We can therefore rewrite Eq. (9) as

$$\text{HBT} = \left\{ 1 + \frac{(\bar{z}_1 - \bar{z}_2)^2}{4\pi^2(\bar{\sigma}_g^2 + \bar{z}_1 \bar{z}_2 / 4\pi^2 \bar{\sigma}_I^2)^2} \right\}^{-1/2}, \qquad (9a)$$

which is approximately a Lorenzian distribution with a width of $2\pi(\bar{\sigma}_g^2 + \bar{z}_1 \bar{z}_2 / 4\pi^2 \bar{\sigma}_I^2)$. Hence, for small $\bar{\sigma}_I^2 / \bar{z}_1 \bar{z}_2$ (the far field), the decrease of HBT with the increase of $(\bar{z}_1 - \bar{z}_2)$ is slow, see Fig. 2. That is to say, the smaller $\bar{\sigma}_I^2 / \bar{z}_1 \bar{z}_2$ is, the easier the HBT is measured experimentally. This is why the HBT experiment is usually done in the far field. However, the HBT experiment can be carried out principally with the near field (or not-far field), because at $(\bar{z}_1 - \bar{z}_2) = 0$ we always have HBT = 1, no matter what is the value of $\bar{\sigma}_I$ (and $\bar{\sigma}_g$). Large $\bar{\sigma}_I^2 / \bar{z}_1 \bar{z}_2$ (near field or not-far field) does not change the physics of the HBT experiment (the intensity fluctuations), but increases the difficulty of the HBT experiment.

For the GI experiment, it follows on substituting Eqs. (4) and (5b) into Eq. (3) and with $z_1 = z_2$ where the image is formed, we obtain the cross correlation function and the intensity

$$\Gamma(u_1,0) = \frac{4\pi G_0 \exp(\frac{i\pi}{\bar{z}_1}\bar{u}_1^2)}{\bar{f}^{1/2}\bar{\xi}^{1/2}} \int d\bar{v} H(\bar{v}) \exp(-\frac{i\pi}{\bar{z}_1}\bar{v}^2)$$
$$\times \exp\{-\frac{4\pi^2[(\bar{\sigma}_g^2 + 2\bar{\sigma}_I^2)\bar{u}_1^2 - 4\bar{\sigma}_I^2\bar{u}_1\bar{v} + (\bar{\sigma}_g^2 + 2\bar{\sigma}_I^2)\bar{v}^2 - i4\pi\bar{\sigma}_g^2\bar{\sigma}_I^2(\bar{v}^2 - \bar{u}_1^2)/\bar{z}_1]}{\bar{\sigma}_g^2\bar{\sigma}_I^2\bar{\xi}}\},$$

(10a)

$$\langle I(u_2=0) \rangle = \frac{4\pi G_0}{\bar{f}\bar{\xi}^{1/2}} \iint d\bar{v}_1 d\bar{v}_2 H(\bar{v}_1) H(\bar{v}_2) \exp(-\frac{i\pi}{\bar{z}_1}\bar{v}_2^2 + \frac{i\pi}{\bar{z}_1}\bar{v}_1^2)$$
$$\times \exp\{-\frac{4\pi^2[(\bar{\sigma}_g^2 + 2\bar{\sigma}_I^2)\bar{v}_1^2 - 4\bar{\sigma}_I^2\bar{v}_1\bar{v}_2 + (\bar{\sigma}_g^2 + 2\bar{\sigma}_I^2)\bar{v}_2^2 - i4\pi\bar{\sigma}_g^2\bar{\sigma}_I^2(\bar{v}_2^2 - \bar{v}_1^2)/\bar{z}_1]}{\bar{\sigma}_g^2\bar{\sigma}_I^2\bar{\xi}}\},$$

(10b)

and $\langle I(\bar{u}_1) \rangle = (4\pi G_0 / \bar{\xi}^{1/2}) \exp[-8\pi^2 \bar{u}_1^2 / \bar{\sigma}_I^2 \bar{\xi}]$, where $\bar{\xi} = 16\pi^2 + (\bar{z}_1^2/\bar{\sigma}_I^2)(4/\bar{\sigma}_g^2 + 1/\bar{\sigma}_I^2)$. With Eqs. (10a) and (10b), we calculate $\text{GI}(\bar{u}_1, \bar{z}_1 = \bar{z}_2) = |\Gamma(\bar{u}_1)|^2 / \langle I(\bar{u}_1) \rangle \langle I(\bar{u}_2=0) \rangle$ numerically. In Fig. 3, we plot the ghost image for a triple slits object (with the width of each slit being $10\lambda$ and the separation between the two slits being $10\lambda$) with $\bar{z}_1 = \bar{z}_2 = 10^5$ for different values of $\bar{\sigma}_I$ and $\bar{\sigma}_g$. Within the three slits, $H(\bar{v}) = 1$, 0.8, and 0.6, respectively, and is zero elsewhere.

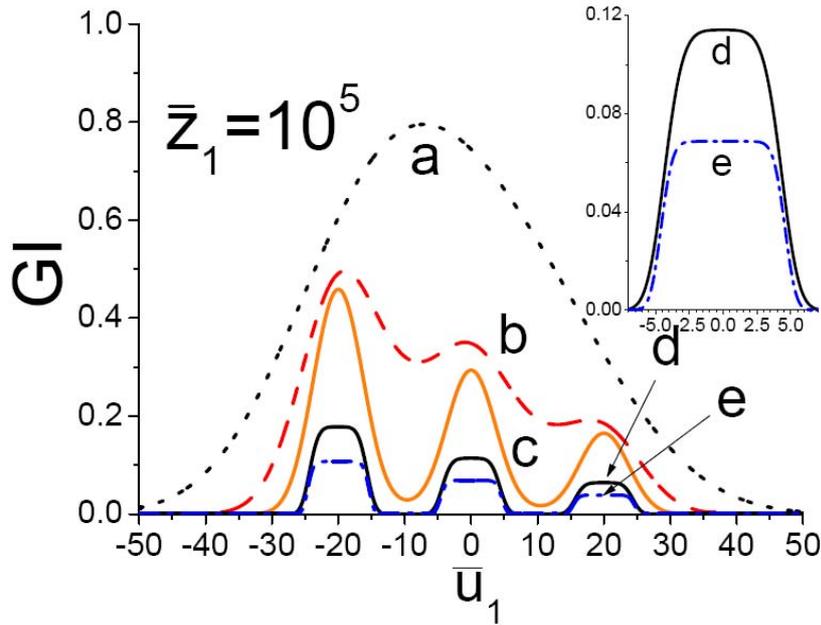

Fig.3, The ghost image of three slits, (a) $\bar{\sigma}_I / \bar{z}_1 = 10^{-2}$, (b) $\bar{\sigma}_I / \bar{z}_1 = 2.5 \times 10^{-2}$, (c) $\bar{\sigma}_I / \bar{z}_1 = 0.2$, with $\bar{\sigma}_g = 4$, (d) $\bar{\sigma}_I / \bar{z}_1 = 0.2$ and $\bar{\sigma}_g = 1$, and (e) $\bar{\sigma}_I / \bar{z}_1 = 0.2$ and $\bar{\sigma}_g = 0.1$.

From Fig. 3, we see that, for small $\bar{\sigma}_I / \bar{z}_1$ (far field) (curves a and b), there is no image. Note that decreasing $\bar{\sigma}_g$ does not help for small $\bar{\sigma}_I / \bar{z}_1$. For a good quality image we need large $\bar{\sigma}_I / \bar{z}_1$ (near field or not-far field) and small $\bar{\sigma}_g$. In curves (d) and (e), we note the formation of the image and the image edge of the middle slit spreads approximately from 3 to 7 and 4 to 6 (see the inset) with visibilities of 12% and 7%, respectively. When we have good quality image, the visibility is low. In addition, large size slits result in low visibility. Therefore, large $\bar{\sigma}_I / \bar{z}_1$ (near field or not-far field) increases the difficulty for detection in both experiments.

For small $\bar{\sigma}_I / \bar{z}_1 \ll 1$ (far field) we still have HBT effect, but no thermal light ghost image. The difference can be explained as follows. In the HBT experiment, the measurement mainly differentiates between two values, HBT = 1 or 0. This corresponds to one bit information. On the other hand, in the ghost image, we need to obtain the information of the whole object and this corresponds to a large amount of bits. The large amount of bits is processed together and one particular bit must not be influenced by other bits. For the curve (c) in Fig. 3, it is hard to say whether the image is formed. If we consider the three slits as three bits, we can conclude: "yes, we have three bits".

Let us consider a very narrow slit for the object located at $\bar{u}_{20}$. The image measurement becomes the determination of one nonzero value at one location (and near by) and zero value at other locations, which is equivalent to the measurement in the HBT experiment to obtain one bit information. Setting $\bar{z}_1, \bar{\sigma}_I \gg \bar{\sigma}_g$ (valid in any experiment) we have

$$\frac{|\Gamma(\bar{u}_1, \bar{u}_{20})|^2}{\langle I(\bar{u}_1) \rangle \langle I(\bar{u}_{20}) \rangle} = \exp[-\frac{(\bar{u}_1 - \bar{u}_{20})^2}{\bar{\sigma}_g^2 + \bar{z}_1^2 / 4\pi^2 \bar{\sigma}_I^2}]. \qquad (11)$$

The image of the very narrow slit is a Gaussian distribution with a width of $\zeta^2 = \{\bar{\sigma}_g^2 + \bar{z}_1^2 / 4\pi^2 \bar{\sigma}_I^2\}$ (the same denominator for the HBT effect). In fact, this is a transverse HBT effect. For small $\bar{\sigma}_I / \bar{z}_1$ ($\ll 1$, far field), the ghost image of the very narrow slit can still be formed but with very bad quality (wide spread). Large $\bar{\sigma}_I/\bar{z}_1$ (= 0.2, not far field) and small $\bar{\sigma}_g$ lead to small width (good image quality), a similar situation for the HBT experiment. In order to have good image, each point at the object should form its own point image with as small as possible spread. This can not be achieved for small $\bar{\sigma}_I/\bar{z}_1$ (far field) even in the limit $\bar{\sigma}_g \to 0$.

From Eq. (9a) we know if $\bar{\sigma}_g \to 0$ and $\bar{\sigma}_I^2/4\pi^2\bar{z}_1\bar{z}_2 = 1$, the width of HBT is one wavelength (very good bunching). Let us consider the theoretical limit $\bar{\sigma}_g \to 0$ and $\bar{\sigma}_I^2/\bar{z}_1\bar{z}_2 \gg 1$. Under this limit, we have HBT = 1 at $(\bar{z}_1 - \bar{z}_2) = 0$, and HBT $\approx 0$ for $(\bar{z}_1 - \bar{z}_2) \neq 0$, which can be considered as perfect bunching; photo-electrons always comes out in pairs. For an easy detection, imperfect bunching is used, so that 1 > HBT > 0 for some value of $(\bar{z}_1 - \bar{z}_2) \neq 0$. For $\bar{\sigma}_g \to 0$ and $\bar{\sigma}_I^2/4\pi^2\bar{z}_1\bar{z}_2 = 1$, one point object will form an image with a size of one wavelength, see Eq. (11). Under the limit ($\bar{\sigma}_g \to 0$ and $\bar{\sigma}_I/\bar{z}_1 \gg 1$) the perfect image will be formed; or we can say the perfect bunching results in the perfect image. It can be proven from Eqs. (10) that under this limit we have $\Gamma(\bar{u}_1, \bar{u}_2 = 0) \propto H(\bar{u}_1)$ [13] with

$$|\Gamma(\bar{u}_1, \bar{u}_2 = 0)|^2 / \langle I(\bar{u}_2 = 0)\rangle\langle I(\bar{u}_1)\rangle = |H(\bar{u}_1)|^2 / \int d\bar{v} |H(\bar{v})|^2, \qquad (12)$$

which means perfect imaging with very low visibility, as the size of any object is much larger than $\lambda$. Large-size source improves the quality of the image, and at the same time reduces the visibility due to the strong background intensity, which comes from the bucket detection triggered by the thermal light passing through different points of the object.

We conclude that the physics behind the thermal light ghost imaging is the intensity fluctuations, the same as for the HBT experiment. The difference between the GI and the HBT experiments is the information that is required to be obtained: large amount for GI and a small amount for HBT (large number of bits versus one bit). In the HBT experiment the far field is used for the purpose of easy detection, while in the GI experiment the near field (not-far field) is used for good quality image at the expense of low visibility.

**Acknowledgement:** This work is supported by HKUST3/06C of HK Government, FRG of HKBU, and NSFC (Contract No. 10604047). The research of MSZ is supported by a grant from Qatar National Research Fund (QNRF).